\documentclass[aps,prl,twocolumn,letterpaper]{revtex4}

\usepackage{amssymb}
\usepackage{amsmath}
\usepackage{graphicx}

\begin{document}

\title{Modeling Connectivity in Terms of Network Activity}

\author{Lucas Antiqueira$^1$}
\author{Francisco Aparecido Rodrigues$^1$}
\author{Luciano da Fontoura Costa$^{1,2,}$}
\affiliation{$^1$Institute of Physics at S\~{a}o Carlos, University of S\~{a}o Paulo, S\~{a}o~Carlos, SP, Brazil \\
%Av. Trabalhador S\~{a}o Carlense~400, PO~Box~369, Postal~Code~13560-970
$^2$National Institute of Science and Technology for Complex Systems, Brazil}

\begin{abstract}
A new complex network model is proposed which is founded on growth
with new connections being established proportionally to the current
dynamical activity of each node, which can be understood as a
generalization of the Barab\'{a}si-Albert static model.  By using
several topological measurements, as well as optimal multivariate
methods (canonical analysis and maximum likelihood decision), we show
that this new model provides, among several other theoretical types of
networks including Watts-Strogatz small-world networks, the greatest
compatibility with three real-world cortical networks.
\end{abstract}

\maketitle

Several models have been developed in order to better understand the
structure and evolution of complex networks, including Erd\H{o}s and
R\'{e}nyi random graph model (ER)~\cite{Erdos1959} and Watts and
Strogatz small-world model (WS)~\cite{Watts1998}. Another widely known
approach is the Barab\'{a}si-Albert (BA) model, developed as a means
to reproduce the scale-free feature observed in many real-world
networks, such as in the World Wide Web (WWW), the power grid and the
actor collaboration network~\cite{Barabasi1999}. Scale-free networks
present a power-law distribution of degrees in the form $P(k)
\sim k^{-\gamma}$, where $k$ is the degree and the exponent
$\gamma$ is system dependent. Such networks are characterized by the
existence of hubs, i.e.\ nodes with particularly high degrees. The BA
model is based on growth and preferential attachment: starting with a
small network of $n_0$ nodes, new nodes are sequentially added and
connected to $m$ other nodes with probability $\Pi(i) =
k_i/\sum_j{k_j}$ ($k_i$ is the degree of node $i$), i.e. nodes with
higher degree have a proportionally higher probability of receiving
new connections~\cite{Barabasi1999}.

Other models of network formation have been proposed as modifications
or generalizations of the BA model, e.g. non-linear preferential
attachment rules~\cite{Krapivsky2000}. Nevertheless, these models tend
to be solely based on the structural features of the growing network,
such as the degree. In the work reported here, we have developed a
preferential attachment model based on a \emph{dynamical} feature of
network nodes: namely, nodes with higher activity have higher
probability of establishing new connections. We use the term
``activity'' to refer to the stationary distribution $\vec{\pi}$ of
frequency of visits to nodes of a random walk, quantified as $\pi_i =
\lim_{t \rightarrow \infty} v_i/t$, where $v_i$ is the number of times
the random walker visited node $i$ after $t$ walking steps. Therefore,
the attachment rule of our model is based on a dynamical process
taking place on the entire network rather than on static, topological
feature of nodes.

It is worth pointing out that the Activity-based Preferential
Attachment model (APA for short) introduced here takes into account
the more general case of \emph{directed} networks, as opposed to the
undirected ones considered in the BA model, so that in- and out-going
edges are assigned for each newly created node. It turns out that if
undirected edges are created in the APA model, the BA model is
reproduced, as a consequence of the fact that the activity is
perfectly correlated with the degree in undirected
networks~\cite{Costa2007b}.  Therefore, the APA model can be
understood as a generalization of the BA model.  However, as in
directed networks the activity is not in general correlated with the
degree, the type of topologies produced by the APA model depends
strongly on the specific distributions of frequencies of visits.  As a
matter of fact, the correlations between in-degree and activity and
between out-degree and activity vary greatly from network to network,
and no exact and general relationship between these quantities has
been discovered yet, except that full correlation between the
frequency of visits and the degree is obtained provided the in-degree
is identical to the out-degree for every node in the
network~\cite{Costa2007b} (observe that the undirected networks are
only a particular case of this more general condition).

Some models based on dynamical features have also been reported in the
literature. Examples include models driven by 
%oscillatory activity~\cite{Gong2003}, 
%fitness of species~\cite{Garlaschelli2007},
Prisoner's Dilemma dynamics~\cite{Poncela2008} and degree of
synchronizability~\cite{Donetti2008}.  Another approach to the
development of connectivity is Hebbian theory~\cite{hebb1949aob},
where two neurons are connected if a neuron repeatedly or persistently
takes part in firing the other neuron. Such a model has been
considered in the study of complex networks~\cite{baryam2004rcn}, but
we are not aware of related growing models.  Indeed, no model based on
preferential attachment taking into account individual node activity
seems to have been proposed yet. Our main motivation to develop such
model is that in many networks the dynamics is more closely related to
the relevance of each node than the degree or other traditional
structural node measurement. For instance, the frequency of visits to
each web page in the WWW is a particularly efficient indicator of page
relevance (viz. Google's PageRank~\cite{Page1998}).
%Indeed, Google's PageRank is successfully based on a random
%walk model to classify a list of web-pages resulting from a 
%query in their web-search mechanism. 
Therefore, a measurement of the dynamics taking 
place in a network constitute a more direct and
reliable indicator of node relevance than just in- or out-degrees,
especially for directed networks, where the activity tends to be 
uncorrelated with the degree.  We have chosen the traditional 
random walk, i.e. the probability of following a link is inversely 
proportional to the out-degree of the current node,
because it is one of the most important models of dynamics in 
physics and in many other fields~\cite{Lovasz1996}.

In the present work, we show that several directed networks
intrinsically underlying dynamical processes, such as the cortical
networks of the cat and the macaque~\cite{scannell1999coc, 
honey2007nsc}, are best reproduced by the APA model
than by other classical complex network models (e.g. WS and BA). Our
methodology is based on the characterization of each network in terms
of a set of network measurements and on the subsequent application of 
canonical variable analysis and Bayesian decision 
theory~\cite{Costa2005a,Duda2000pc}.

A directed network of $N$ nodes can be completely represented by the
adjacency matrix $A$ of order $N \times N$, whose position $A(j,i)$ is
equal to 1 if and only there exists a directed edge pointing from node
$i$ to node $j$ (otherwise, $A(j,i) = 0$). The in-degree of node $i$,
i.e. the number of connections it receives, is equal to $k_i^{in} =
\sum_j{A(i,j)}$, and its out-degree, which is equal to the total
number of edges leaving it, is given as $k_i^{out} =
\sum_j{A(j,i)}$~\cite{Costa2005a}. The dynamics of a random walk is
entirely determined by the stochastic (or transition) matrix $P$ with
elements $P(j,i) = 1/k_i^{out}$, i.e. the probability of the walker
visiting node $j$ at step $t+1$ after being at node $i$ at step $t$ is
equal to $P(j,i)$.  Notice that $A(j,i)=0$ implies $P(j,i)=0$. In
other words, the next step of the walker depends only on its current
state (Markov Chain). The stationary, or steady-state, distribution of
probabilities of being at each node, i.e. $\vec{\pi}$, can be obtained
by solving $P\vec{\pi} = \vec{\pi}$.  In particular, $\vec{\pi}$ is
the eigenvector associated with the eigenvalue 1 of
$P$~\cite{Lovasz1996}. We define $\sum_i{\pi_i} = 1$ for proper
statistical normalization. For this distribution to be unique, $P$
needs to be irreducible, i.e. the network must be strongly connected,
which happens when every node can reach every other node in the
network through a finite path. For undirected networks (when $A$ is a
symmetric matrix), the stationary distribution of a random walk can be
directly obtained from the degree distribution as follows $\pi_i = k_i
/ \sum_j{k_j}$, whereas for directed networks this perfect correlation
only happens when, for every node $i$, $k_i^{in} = k_i^{out}$.
%\footnote{More generally, the perfect correlation in weighted networks 
%happens when $s_i^{in} = s_i^{out}$ for every node $i$, where $s_i^{in}$ 
%and $s_i^{out}$ are the in- and out-strengths of the node $i$,
%respectively~\cite{Costa2007b}.}

The growth of the APA model starts at $T=0$ with a random directed
network of $n_0$ nodes and initial walker distribution
$\vec{\pi}^0$. At each subsequent time step $T$ ($T \leq N-n_0$) a new
node $i$ is added to the network and directed edges are independently
created from $m$ older nodes ($m \ll n_0$) to the current node $i$
(in-edges) and from node $i$ to $m$ other nodes (out-edges). The
general idea behind this model is that a new node would want to
establish in-edges with highly active nodes in order to receive a
considerable share of activity from the outset of its
lifetime. Therefore, in-edges are created between node $i$ and other
nodes following the preferential attachment probability $\Pi(i) =
\pi_i^{T-1}$. Because there is no out-edge attachment rule that
could intuitively increase the activity of a new node, we have
proposed two approaches: a uniform rule $\Pi(i) = 1/N$, and the
preferential rule already used for in-edges. Therefore, these two
approaches have divided APA into two model variations: the original
APA, that considers the preferential attachment for both edge 
directions, and APA', that takes into account only the in-edges
in the preferential attachment rule.

\begin{figure}[htb]
	\centering
		\includegraphics[width=0.7\columnwidth]{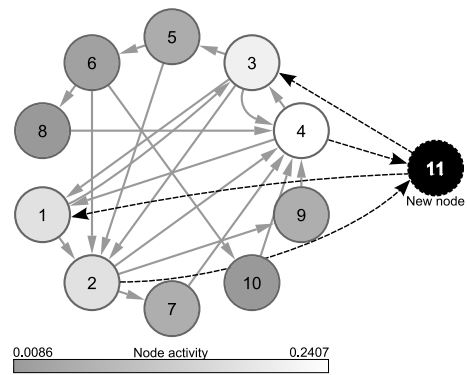}
	\caption{A step $T$ along the growth of a network by using the
	APA model. Ten nodes already exist in the network,
	whose activities $\pi_i^{T-1}$ are depicted by gray levels, 
	and a new node tends to make connections with highly active nodes.
	%In this example, a new node establishes $m=2$
	%connections in both directions, with higher probability of
	%choosing highly active nodes.
	}  
	\label{fig:fig1}
\end{figure}

Figure~\ref{fig:fig1} illustrates a small network being constructed by
the APA approach with $m=2$, where a new node tends to be connected
with highly active nodes regarding both edge directions.  The
stationary frequencies of visits of each node are shown by
gray-levels.  Notice that for both APA and APA' networks, 
the average in- and out-degrees are $\left\langle k^{in} 
\right\rangle =\left\langle k^{out} \right\rangle = 2m$. 
The BA model is exactly reproduced by considering 
the APA model with undirected edges, since $\Pi(i) =
\pi_i^{T-1} = k_i / \sum_j{k_j}$ for every undirected and connected
network. Consequently, the APA model can be understood as a
generalization of the BA model.

\begin{figure*}[htb]
	\centering
		\includegraphics[width=0.8\linewidth]{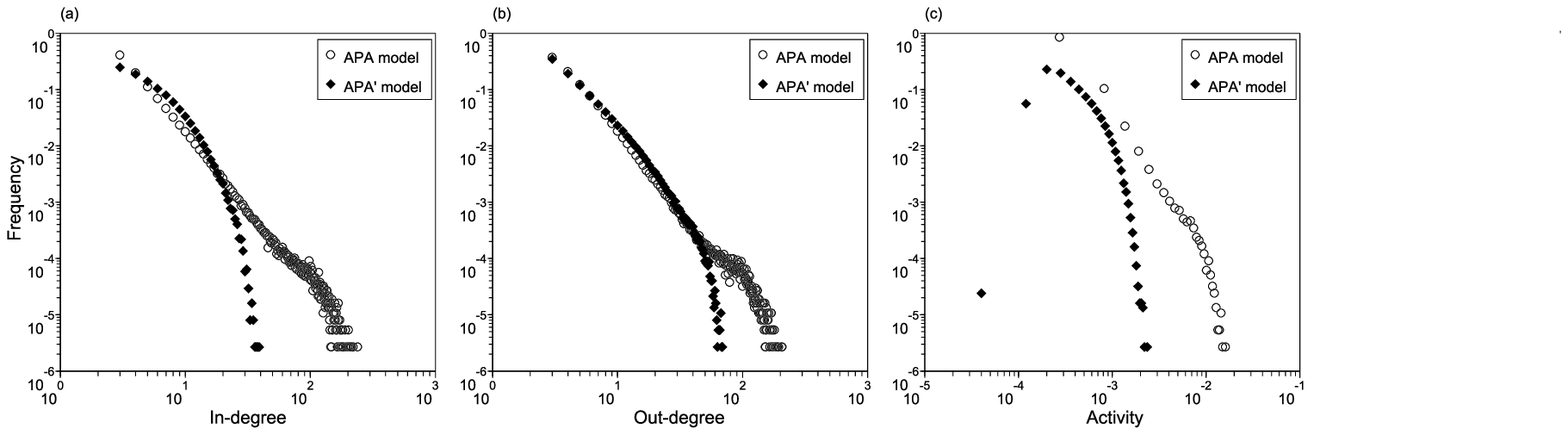}
	\caption{Average (a) in-degree, (b) out-degree and (c)
	activity distributions of models APA and APA'. Standard
	deviations are too small for visualization.}  
	\label{fig:fig2}
\end{figure*}

Computer simulations were performed in order to characterize APA and
APA' models, where 150 realizations of each model with
$N=2500$ and $\left\langle k^{in} \right\rangle =\left\langle k^{out}
\right\rangle = 6$ were generated, and the respective average results
are shown in Figure~\ref{fig:fig2} and Table~\ref{tab:table1}. The in-
and out-degree distributions are included in Figure~\ref{fig:fig2}, as
well as the activity distribution. The APA model yields power-law
degree distributions for both edge directions, although these
distributions have a small deviation from a power-law when considering
high degrees. Nevertheless, the obtained distributions clearly show
the existence of hubs of in- and out-degree in the APA model. When
considering APA' networks, these distributions greatly deviate from a
power-law, with steep decays for high degrees that prevent the
occurrence of nodes as highly connected as in the APA networks. The
activity distributions reveal that APA networks are also scale-free
with this respect, whereas APA' again deviates from a power-law, not
showing hubs of activity like in the APA model. Networks with 
different sizes and average degrees were also analyzed, yielding 
similar results.

Table~\ref{tab:table1} shows the values obtained for a set of
measurements calculated for APA and APA' (each model with 
150 realizations, $N=2500$ and $\left\langle k^{in} 
\right\rangle = \left\langle k^{out} \right\rangle = 6$). 
The employed measurements are~\cite{Costa2005a,Garlaschelli2004} 
(i)~clustering coefficient $C$, (ii)~length of shortest paths 
$\ell$, (iii)~assortativity between in- and out-degrees 
($r^{ii}$ and $r^{oo}$, respectively) and (iv)~reciprocity 
$\rho$, where averages were taken for each network model.
%The clustering coefficient $C$ is a well-known measurement 
%that quantifies the level of connectivity among the 
%neighbors of nodes~\cite{Watts1998}. 
%The average length of shortest paths takes into account the minimum 
%number of edges (on average) necessary to reach 
%a node $j$ after starting at a node $i$~\cite{Watts1998}. The
%assortativity $r$ of a network is a measure of the correlation
%(Pearson coefficient) between the degrees of pairs of interconnected
%nodes. It indicates, for instance, if highly connected nodes tend to
%be linked to other well-connected nodes~\cite{Newman2002}. 
%The degrees are uncorrelated when $|r| \approx 0$ and 
%correlated when $|r| \approx 1$ (in this case, the sign 
%of $r$ gives the sign of the correlation). 
%Since there are in- and out-degrees, we used two types
%of assortativity: $r^{ii}$ and $r^{oo}$, correlating the in- and
%out-degrees, respectively.
The first three measurements are well-known~\cite{Costa2005a}, 
while the reciprocity quantifies to what an 
extent a directed network contains symmetric links, i.e. edges that 
connect pairs of nodes at both directions~\cite{Garlaschelli2004}. 
%It is given by 
%$\rho = (L^{\leftrightarrow}/L - \bar{a})/(1 - \bar{a})$, where 
%$L^{\leftrightarrow}$ is the number of reciprocal edges in a 
%network, $L$ is its total number of edges and 
%$\bar{a} = L/(N(N-1))$ is its link density. 
%Reciprocal ($\rho>0$) or antireciprocal ($\rho<0$) networks are, 
%respectively, those with more or less symmetric edges than what 
%would be expected by chance (areciprocal networks have $\rho = 0$). 
%The maximum reciprocity is $\rho_{max} = 1$ and the minimum is 
%$\rho_{min} = -\bar{a}/(1 - \bar{a})$.
We also computed the minimum reciprocity $\rho_{min}$, which depends 
on the specific edge density of a network. 

\begin{table}[htb]
	\centering
	\caption{Average and standard deviations of the clustering
	coefficient ($C$), length of shortest paths ($\ell$), in- and
	out-assortativities ($r^{ii}$ and $r^{oo}$) and reciprocity
	($\rho$ and $\rho_{min}$), for models APA and APA'.}  
	\small
	\begin{tabular}{l@{\hspace{20pt}}r@{\hspace{20pt}}r}
		\hline \hline
		& APA model & APA' model \\
		\hline
		$C$          &  $0.0306 \pm 0.0023$   & $0.0090 \pm 0.0004$ \\
		$\ell$       &  $3.7453 \pm 0.0240$   & $4.2433 \pm 0.0075$ \\
		$r^{ii}$     & $-0.0584 \pm 0.0063$   & $0.1512 \pm 0.0095$ \\
		$r^{oo}$     & $-0.0500 \pm 0.0071$   & $0.2094 \pm 0.0107$ \\
		$\rho$       &  $0.0123 \pm 0.0015$   & $0.0063 \pm 0.0009$ \\
		$\rho_{min}$ & $-0.0024 \pm 10^{-6}$ & $-0.0024 \pm 10^{-6}$ \\
		\hline \hline
	\end{tabular}
	\label{tab:table1}
\end{table}

The results included in Table~\ref{tab:table1} show that both APA
and APA' networks have particularly low clustering coefficients
(specially in APA') coexisting with short path lengths (around~4),
therefore resembling the BA model. The assortativity values show no
correlation between degrees in the APA model (like in the BA model),
whereas slightly positive correlations appear in APA'. Reciprocity
results indicate that APA and APA' networks are areciprocal ($\rho
\approx 0$), which means that edges in these networks do not tend to
be symmetric (reciprocity results are also particularly close
to the respective minimums). Similar characteristics were observed 
in networks of different sizes and densities (results not shown).

In order to accurately compare the models APA and APA' with other
models, and also to classify real-world networks with reference to a
set of putative models, it is necessary to consider a larger number of
network measurements~\cite{Costa2005a}. 
%In fact, accurate characterization and classification of network 
%structures can be done by using a large set of measurements~\cite{Costa2005a}.  
In the current letter, we adopted a set of nine measurements to quantify the
topological properties of complex networks: (i)~average node degree,
(ii)~average clustering coefficient, (iii)~average
shortest path length, (iv)~in- and (v)~out-assortativity 
coefficients, (vi)~central point dominance, (vii)~average 
betweenness centrality, (viii)~hierarchical clustering
coefficient and (ix)~hierarchical convergence ratio. All 
these measurements are discussed in~\cite{Costa2005a}.

The classification of real-world networks involves the consideration
of different network models, each one generating specific types of
topologies. In this way, we took into account the following models (as
well as APA and APA' models): (i)~Erd\H{o}s-R\'{e}nyi random graph
(ER)~\cite{Erdos1959}, which generates networks with random placement
of connections, (ii)~small-world model of Watts and Strogatz~(WS),
which produces networks whose structures lie between a regular and a
random network~\cite{Watts1998}, (iii)~Barab\'{a}si-Albert scale-free
model~(BA), which constructs networks having a power-law degree
distribution~\cite{Barabasi1999}, and (iv)~a geographical model~(GG),
where nodes next to each other in a given metric space are more likely
to be connected by an edge~\cite{Costa2005a}.

Since the network classification requires a large set of model
realizations in order to minimize statistical fluctuations, 
we applied canonical variable analysis to
reduce the dimensionality of the $M$-dimensional measurement space
while maximizing the separation between the network models.
This multivariate statistical technique is an extension of principal
component analysis and allows optimal projections of a set of
measurements so as to obtain the reduction of the dimensionality of
the original space while maximizing the separation between the known
categories (i.e. network models)~\cite{Campbell81}. The computation of
the canonical variable analysis is based on the so-called inter- and
intra-class matrices, as well as on the diagonalization of the product
between the inverse of the intra-class matrix and the inter-class
matrix. The selection of the $d$ eigenvectors corresponding to the
highest absolute eigenvalues of this matrix product allows the
projection of the measurements into a $d$-dimensional space ($d \leq M$). In the current
work, we considered $d=2$.

The projection is then performed by calculating the inner products
between the original feature (network measurement) vectors and the two
eigenvectors corresponding to the highest eigenvalues.  After the
projection, we estimated the probability density of the projected
points into the two dimensional space for each class, by
considering the non-parametric method called Parzen
windows~\cite{Duda2000pc}.  This method starts by representing each
point in the projection as a Dirac's delta function.  These deltas are
then convolved with a normalized Gaussian function, yielding the
estimated distributions, which are then considered for the
classification of a given real-world network.  Notice that every model
realization has the same number of vertices and approximately the same
average degree of the respective real-world network.  Maximum
likelihood decision theory was then applied in order to classify each
real-world network by associating it to the model that results in the
largest overall probability~\cite{Duda2000pc,Costa2005a}.
Observe that equiprobability of the mass probabilities is guaranteed
by using the same number of realizations for each category.

\begin{figure*}[htb]
	\centering
    	\includegraphics[width=0.6\columnwidth]{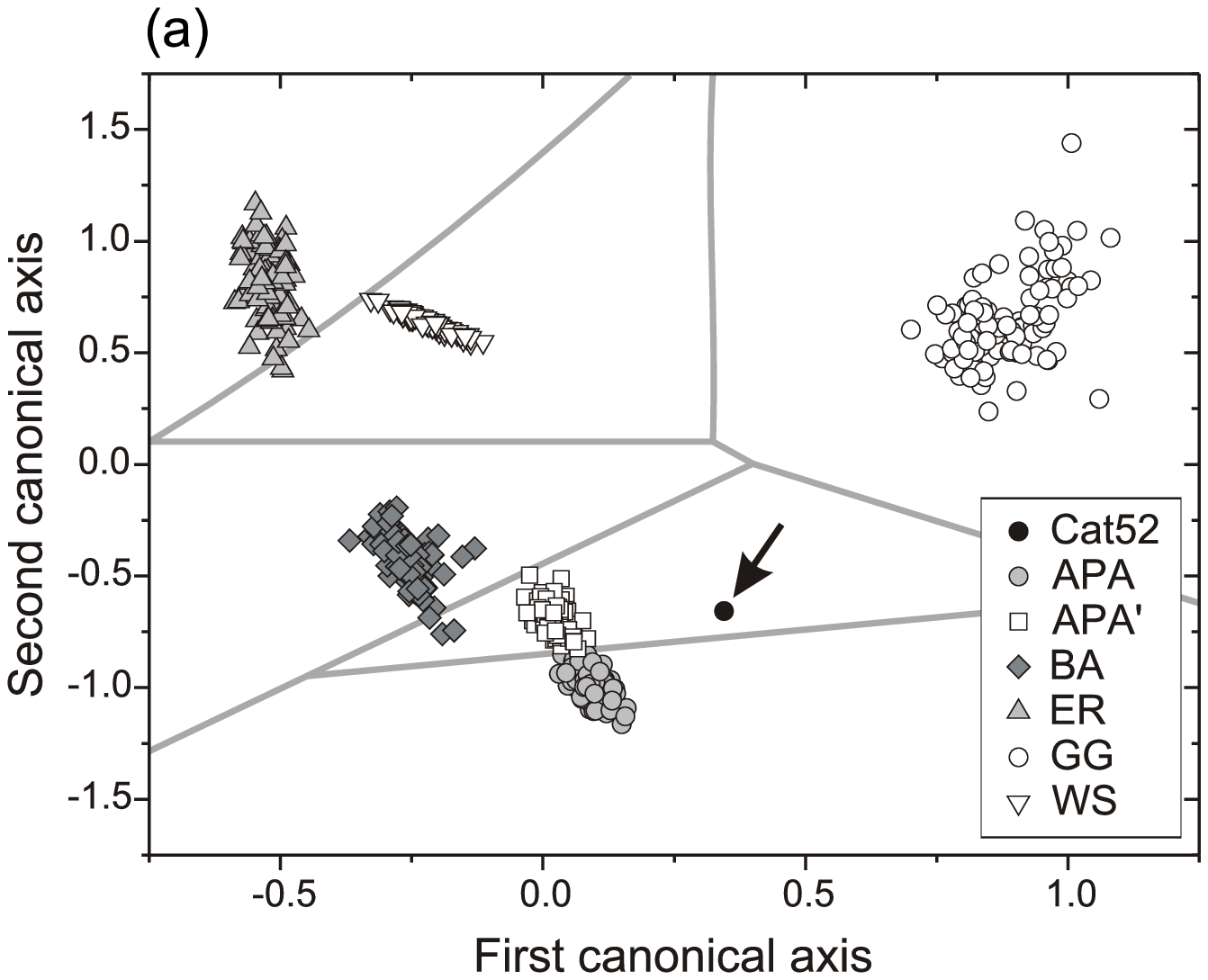}~~~~\includegraphics[width=0.6\columnwidth]{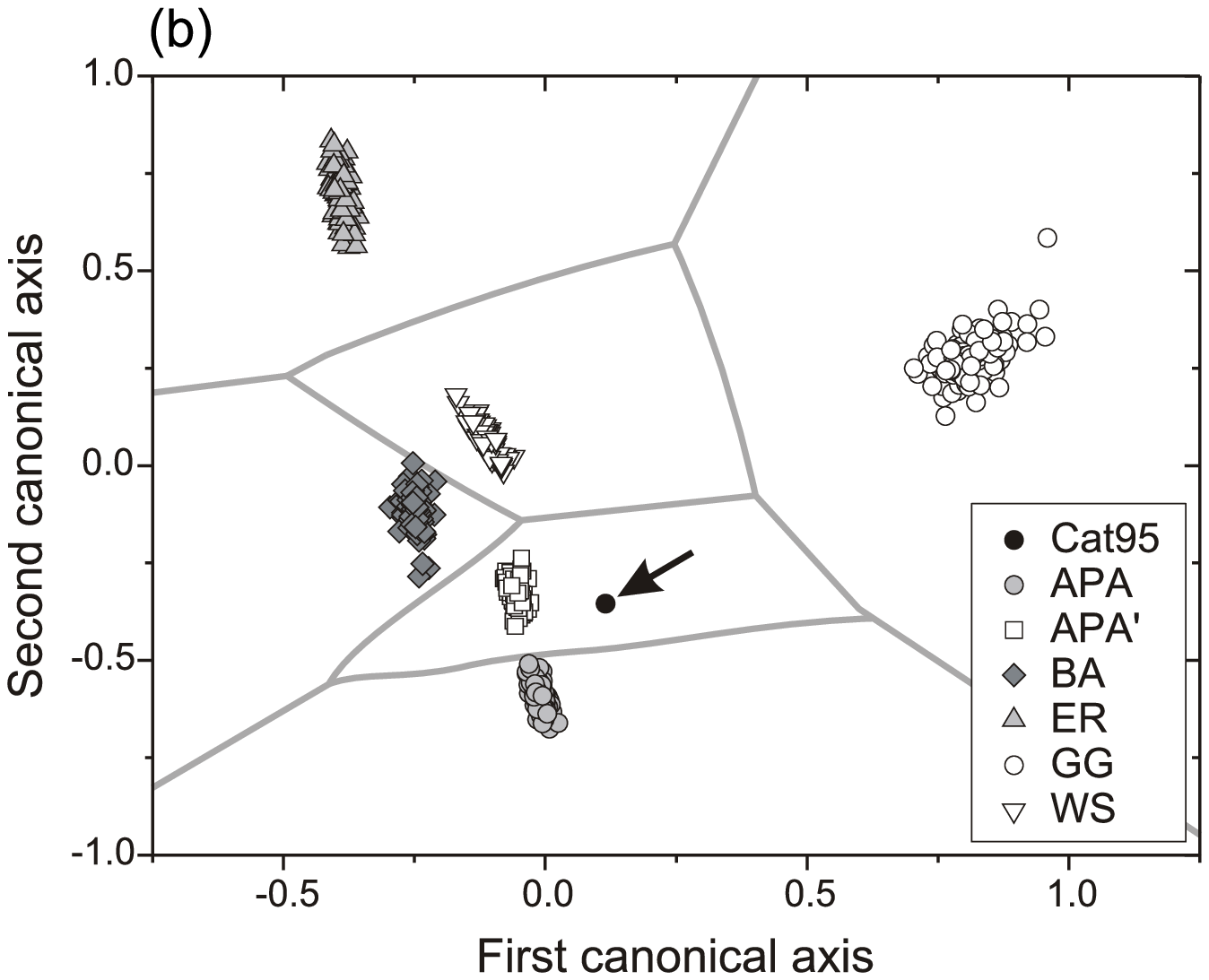} \\
		\includegraphics[width=0.6\columnwidth]{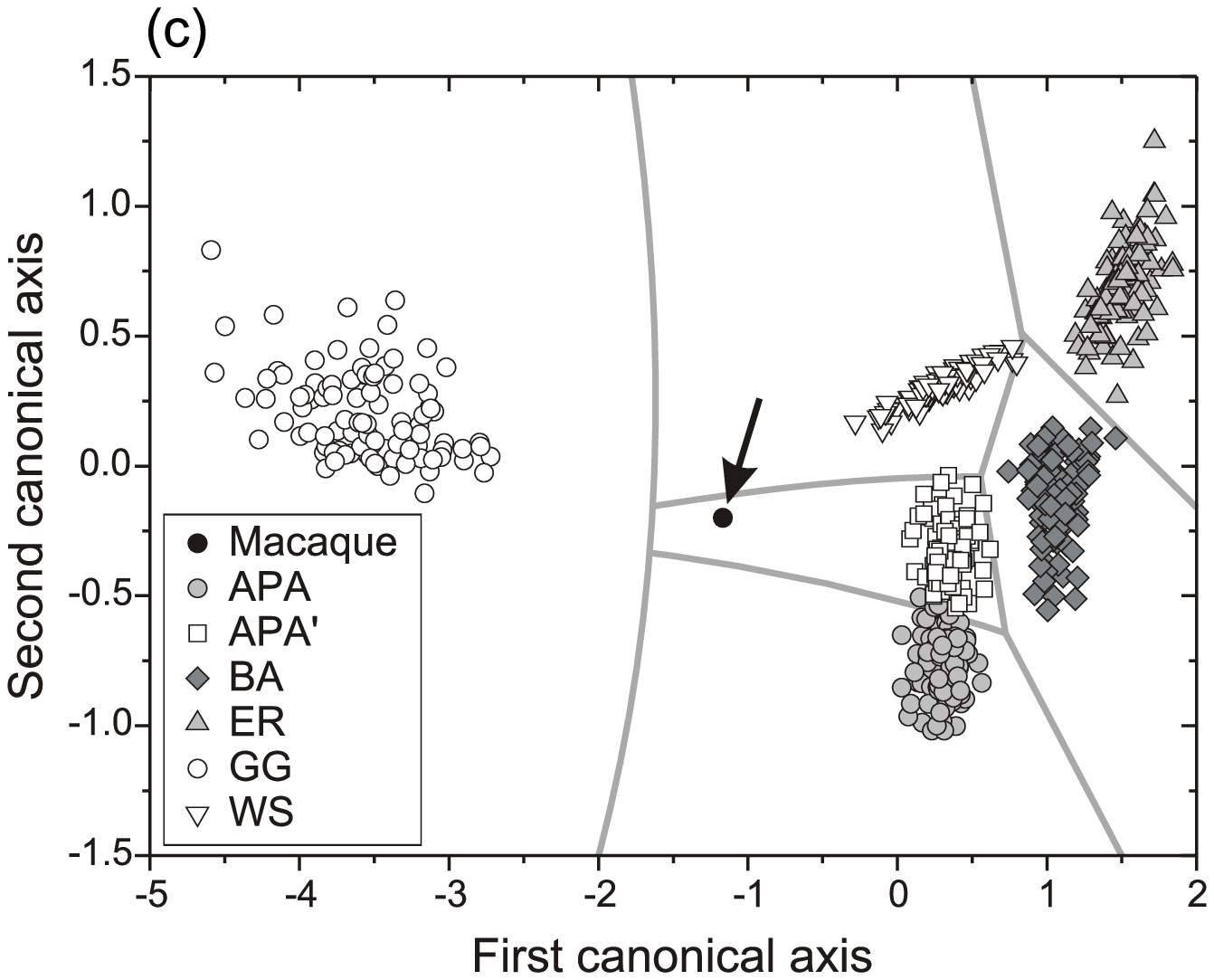}~~~~\includegraphics[width=0.6\columnwidth]{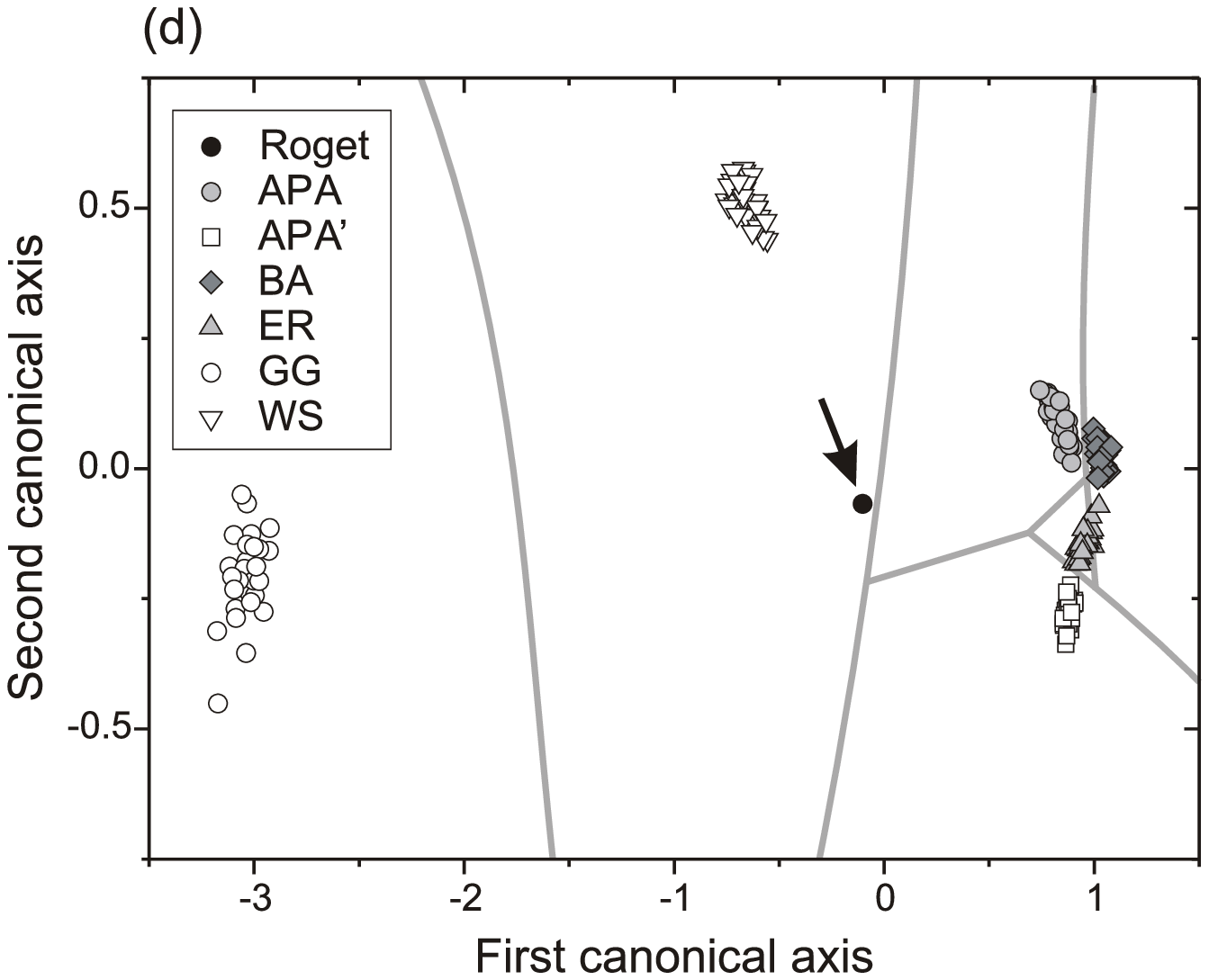}
	\caption{Classification of real-world networks: (a)~cat
	cortex including 52 cortical areas, (b)~cat cortex containing
	all cortical and thalamic areas, (c)~macaque cortical
	network, and (d)~Roget thesaurus network.
    Each classified network is indicated by an arrow and a black circle.}
	\label{fig:fig3}
\end{figure*}

The classification methodology was applied to three cortical networks,
namely (i)~cat cortex containing $N=52$ cortical
areas~\cite{scannell1999coc}, (ii)~cat cortex including
all cortical and thalamic areas ($N=95$)~\cite{scannell1999coc} 
and (iii)~macaque large scale cortical network, including
visual and sensorimotor areas ($N=47$)~\cite{honey2007nsc}; to two 
food webs, (i)~Canton (a pasture grassland in New Zealand,
$N=110$)~\cite{jaarsma1998cfw} and (ii)~Kyeburn (a
tussock grassland in New Zealand, $N=98$)~\cite{jaarsma1998cfw}; 
and to the Roget thesaurus network
($N=1022$)~\cite{dejesusholanda2004tcn}.  For short, the first two
networks are called here ``cat52'' and ``cat95'',
respectively. Figure~\ref{fig:fig3} presents the classification of
all three cortical networks and the Roget thesaurus, taking into
account 100 realizations of each network model. All cortical networks
fall inside the APA' region in the canonical projections, which means
that the APA' model is the most likely in those cases. On the other
hand, in the case of the Roget thesaurus network and the food webs,
the respective networks have been classified as small-world, as could
be expected~\cite{dejesusholanda2004tcn, montoya2002swp}.

Therefore, the proposed activity-based preferential attachment model
(APA') revealed to be compatible with cortical networks. Nevertheless,
it has been claimed that brain networks can be well modeled by WS
networks (e.g.\ \cite{sporns2004swc, hilgetag2004coc,
bassett2006swb}), which is not supported by the results of
the current work. Furthermore, the ``cat95'' network was previously
analyzed using the stationary random walk distribution to simulate
cortical activation~\cite{Costa2006}, and the out-degrees were found
to be highly correlated with random walk activity.

The obtained results suggest that the brain is organized in order to
favor the connections with a small number of highly active regions.  
In fact, since this mechanism is a generalization of the
well-established preferential attachment model, we observed that the
connections in the brain would not be guided by structural aspects
(i.e.\ the connectivity), but rather by dynamics (i.e.\ frequency of
visits by random walks). This process can be related to the Hebbian
theory~\cite{hebb1949aob}, since nodes that receive walks more
frequently tend to receive repeated and persistent stimulation and
therefore establish a large number of connections.

All in all, the current work described a new model of complex network
which is founded on a growth mechanism favoring attachments
proportional to the current activity of each node. This approach 
can also be understood as a generalization of the BA model.
%to incorporate preferential attachment rules proportional to the 
%dynamical activity at each node.  
As such, the proposed approach is intrinsically suited
for modeling complex systems whose connectivity is determined by the
respective dynamics. An effective combination of estimation of several
topological measurements, as well as the optimal methods of canonical
analysis and maximum likelihood decision theory, paved the way to a
sound comparison between real-world networks and several putative
theoretical models, namely APA, APA', ER, WS, BA and GG.  The proposed
model turned out to exhibit a remarkable compatibility with three
real-world cortical networks.
%even exhibiting particularly low clustering coefficients.  
Future developments include the application to modeling neuronal 
networks where each node represents a neuron.  It would 
also be interesting to consider attachment rules founded 
on transient, rather than stationary dynamics.

\begin{acknowledgments}
L.~da F.~Costa is grateful to FAPESP (05/00587-5) and CNPq
(301303/06-1) for financial support. F.~A. Rodrigues
thanks FAPESP grant 07/50633-9 and L.~Antiqueira
FAPESP grant 06/61743-7.
\end{acknowledgments}

\bibliographystyle{apsrev}
\bibliography{dynamics_model}

\end{document}